\documentclass[%
 aip,
 amsmath,amssymb,
 reprint,%
]{revtex4-1}

\usepackage{graphicx}% Include figure files
\usepackage{dcolumn}% Align table columns on decimal point
\usepackage{bm}% bold math
\usepackage[utf8]{inputenc}
\usepackage[T1]{fontenc}
\usepackage{mathptmx}
\usepackage{siunitx}
\usepackage{xcolor}

\newcommand{\beq}{\begin{equation}}
\newcommand{\eeq}{\end{equation}}
\newcommand{\barr}{\begin{eqnarray}}
\newcommand{\earr}{\end{eqnarray}}
\newcommand{\bseq}{\begin{subequations}}
\newcommand{\eseq}{\end{subequations}}
\newcommand{\ket}[1]{|#1\rangle}

\newcommand{\expectation}[3]{\langle #1|#2|#3\rangle}

\newcommand{\vett}[1]{\textbf{#1}}
\newcommand{\uvett}[1]{\hat{\textbf{#1}}}

\begin{document}

\preprint{AIP/123-QED}

\title{Generation of Half-Integer Harmonics and Efficient THz-to-Visible Frequency Conversion in Strained Graphene}
% Force line breaks with \\

\author{Marco Ornigotti}
 \email{marco.ornigotti@tuni.fi}
\affiliation{Faculty of Engineering and Natural Sciences, Photonics, Tampere University, Tampere, FI-33720 Finland}

\author{Luca Ornigotti}%
\altaffiliation[Also at ]{Quantum Optics, Quantum Nanophysics and Quantum Information
Faculty of Physics, University of Vienna, Boltzmanngasse 5 1090, Vienna, Austria}
\affiliation{Department of Optics, Palack\'y University, 17. listopadu 1192/12,  711 46 Olomouc, Czech Republic}

\author{Fabio Biancalana}
\affiliation{School of Engineering and Physical Sciences, Heriot-Watt University, Edinburgh, UK}%

\date{\today}% It is always \today, today,
             %  but any date may be explicitly specified

\begin{abstract}
We study the generation of harmonics from graphene under the influence of an artificial magnetic field, generated via bending of a graphene flake. We show how the Landau level structure induced by the pseudomagnetic field breaks the centrosymmetry of graphene, thus allowing the generation of even harmonics. We also show, that depending on the impinging pulse duration, the nonlinear signal does not only contain the integer harmonics of the impinging pulse, but also its half-integer ones, due to the peculiar square-root-like nature of Landau levels in graphene.
\end{abstract}

\maketitle

\section{introduction}
Gauge fields are ubiquitous in Nature, and regulate the dynamics of several different fields of physics. Perhaps the most common example of a gauge field is the electromagnetic field, whose quantum, the photon, mediates the interaction between charged matter \cite{ref1}. Besides electrodynamics, gauge fields make their appearance in the Standard Model of particle physics, for example, as a unified way to describe the interaction of matter with the fundamental forces of Nature, excluding gravity\cite{ref2}. In quantum mechanics, the gauge invariance of the electromagnetic field is at the core of the celebrated Aharonov-Bohm effect \cite{ref3}, which paved the way for a deeper understanding of gauge fields in terms of fibre bundles \cite{ref4,ref5,ref6}. Moreover, non-Abelian gauge theories are also an essential ingredient to understand geometrical and Berry phases \cite{ref7,ref8}. In condensed matter physics, gauge fields play a crucial role in understanding long-range interaction, and the emergence of collective phenomena, such as the appearance of Abrikosov vortices \cite{ref9} or topological states of matter \cite{ref5,ref10,ref11}.

In recent years, artificial (or synthetic) gauge fields (AGF) have started to attract considerable attention. Contrary to gauge fields, which arise from \emph{real} fields or geometric connections, AGFs can be tuned ad-hoc by acting on a physical system in a certain, predetermined and tunable way. Typically, the physical mechanism generating AGFs is also different, in nature, than the one generating the actual gauge field it mimics. For example, an artificial magnetic field is experiences by cold atoms in a rotating frame, due to the mathematical equivalence between the Coriolis force and the Lorentz force \cite{ref13}, or for light propagating in a waveguide with a twisted propagation direction, due to the topological equivalence between the twisting and an actual magnetic field applied to the waveguide \cite{ref14}. The advent of AGFs paved the way for completely new research fields, which merge together aspects from topology and set theory with different concepts and methods from various physical disciplines, such as topological mechanics \cite{ref15}, topological condensed matter physics \cite{ref16}, topological atomic physics \cite{ref17, ref18, ref19, ref20}, and topological photonics \cite{ref21,ref22}, where both concepts from lattice gauge field theories \cite{ref23} and condensed matter physics \cite{ref16} were used to create novel ways to control the propagation and coupling of light in wave-guiding structures, such as photonic topological insulators \cite{ref24,ref24bis, ref25}, non-Abelian-like dynamics in engineered waveguide lattices \cite{ref26}, AGF switching using the angular momentum of light \cite{ref27}, and topological protection \cite{ref28,ref29,ref30,ref31}.

Aside from photonics, another platform that in the last decade represented a rich playground for testing the effects of different classes of AGFs has been graphene.  Since its experimental discovery in 2004 by Novoselov and Geim \cite{ref32}, graphene has in fact attracted considerable attention not only for its peculiar band structure \cite{ref33}, its anomalous quantum Hall effect \cite{ref34} or its minimal conductivity \cite{ref35} and universal absorbance \cite{ref35a}, but it also represented an unexpected connection between condensed matter physics and gauge field theory. The application of strain, stress, or bending on graphene, in fact, gives rise to AGFs in the form of effective electric and magnetic fields \cite{ref36,ref37,ref38,ref39}. Moreover, out-of-plane bending of graphene flakes is analog, for the electrons in graphene, to consider their evolution on a curved background, under the action of gravity \cite{ref40,ref41}. 

Graphene also represents a very interesting platform for photonics, mainly for its very large nonlinear response, compared to bulk materials. Recent theoretical \cite{extra1,extra7} and experimental \cite{extra8,extra9,extra10} results, in fact, estimate the bulk-equivalent third-order nonlinear susceptibility of graphene to go from $\chi_{3D}^{(3)}\simeq 10^{-15}$ $m^2/V^2$ in the visible region \cite{extra8} up to $\chi_{3D}^{(3)}\simeq10^{-9}$ $m^2/V^2$ in the terahertz region \cite{extra9}. For comparison, the bulk third-order nonlinearity of a typical silica glass is of the order of $\chi_{crystal}^{(3)}\simeq 10^{-23}$ $m^2/V^2$ (see Ref. \cite{extra11}). Graphene, therefore, possesses an extremely high nonlinear response, several orders of magnitude higher than that of a normal nonlinear material . A typical way to obtain such results experimentally is by means of the Z-scan technique, which is a standard characterisation method for the nonlinear response of bulk materials, and can be applied for graphene as well under certain assumptions (see Ref. \cite{extra10} for details). The results of such measurement leads naturally to a 3D susceptibility, but in case of graphene, these numbers should be taken with a grain of salt, since the notion of bulk susceptibility for monolayer graphene does not make much sense (graphene is , indeed, a true 2D material), and they should be used only for comparison with other materials \cite{extra17}.

The nonlinear response of graphene in the limit of strong magnetic field (i.e., when the magnetic length is much smaller than the wavelength of the impinging electromagnetic pulse) has also been estimated theoretically \cite{extra12} to be inversely proportional to the external magnetic field, i.e., $\chi_{3D}^{(3)}\simeq 5\times10^{-9}/B(T)$ $m^2/V^2$. In the same work \cite{extra12}, moreover, the maximum intensity of the nonlinear signal has also been estimated to be linearly proportional to the applied magnetic field, i.e., $I_3^{(max)}\simeq B$. From these results we see how controlling the magnetic field results in an overall increase in the intensity of the nonlinear signal. This feature, combined with the possibility of creating pseudomagnetic fields through AGFs, could be exploited to reduce the necessary pump intensity to trigger nonlinear phenomena in graphene, and might lead, in the future, to a new generation of integrated nonlinear devices.

Amidst the vast literature on AGFs in graphene, some works have been focusing the attention on their effects on the interaction of graphene with external electromagnetic fields, pointing out how a constant external electric field can drastically modify the arrangement of Landau levels and edge states of bent graphene flakes, introducing a squeezing of the Landau states in graphene \cite{ref42}, while a time-dependent electric field can induce modulations of the angular momentum transfer between light and graphene \cite{ref43}, or induce coherent population transfer between different Landau levels of the valence and conduction band of graphene \cite{ref44}. To the best of our knowledge, however, the influence of AGFs on the nonlinear response of graphene interacting with an ultrashort laser pulse has not been investigated yet. 

In this work, therefore, we present a comprehensive analysis of the interaction of ultrashort light pulses with graphene, in the presence of an AGF. In particular, we consider the case of strained graphene, as presented in Ref. \cite{ref39}, which implements an effective constant, uniform, magnetic field orthogonal to the graphene plane. The presence of an AGF induces Landau levels in both the valence and conduction band of graphene, thus introducing selection rules on the impinging pulse polarisation. Under these assumption, we study different interaction configurations, for impinging pulses of different resonant frequencies, and we show, that the nonlinear signal produced by electrons near the Dirac point of graphene differ substantially from the case of unbent graphene, since the presence of the AGF breaks the symmetry of graphene, thus allowing the appearance of even harmonics in the nonlinear optical response. 

This work is organised as follows: in Sect. \ref{section2}, we briefly review how bending graphene introduces an artificial gauge field, and what are the consequences of that for an electron in graphene.  Then, we present the theoretical framework necessary to describe the interaction of ultrashort pulses with graphene in the presence of AGF in Sect. \ref{section3}. Section \ref{section4} is dedicated to the main results of our work, i.e., the calculation and discussion of the nonlinear response of bent graphene to an external ultrashort pulse. Finally, conclusions and future prospects are discussed in Sect. \ref{section5}.
\section{Artificial Gauge Field from Bending}\label{section2}
When a mechanical strain is applied to graphene, the immediate result is the appearance of an artificial gauge field, whose spatial distribution and orientation depends on the nature of the strain applied \cite{ref46}. The general expression for the induced AGF in the absence of out-of-plane modulations can be then written as \cite{ref39}
\beq
\vett{A}^{(s)}=\pm\frac{s\beta}{a}\left(u_{xx}-u_{yy}\right)\uvett{x}\mp\frac{2s\beta}{a}u_{xy}\uvett{y},
\eeq
\begin{figure}[t!]
\begin{center}
\includegraphics[width=0.5\textwidth]{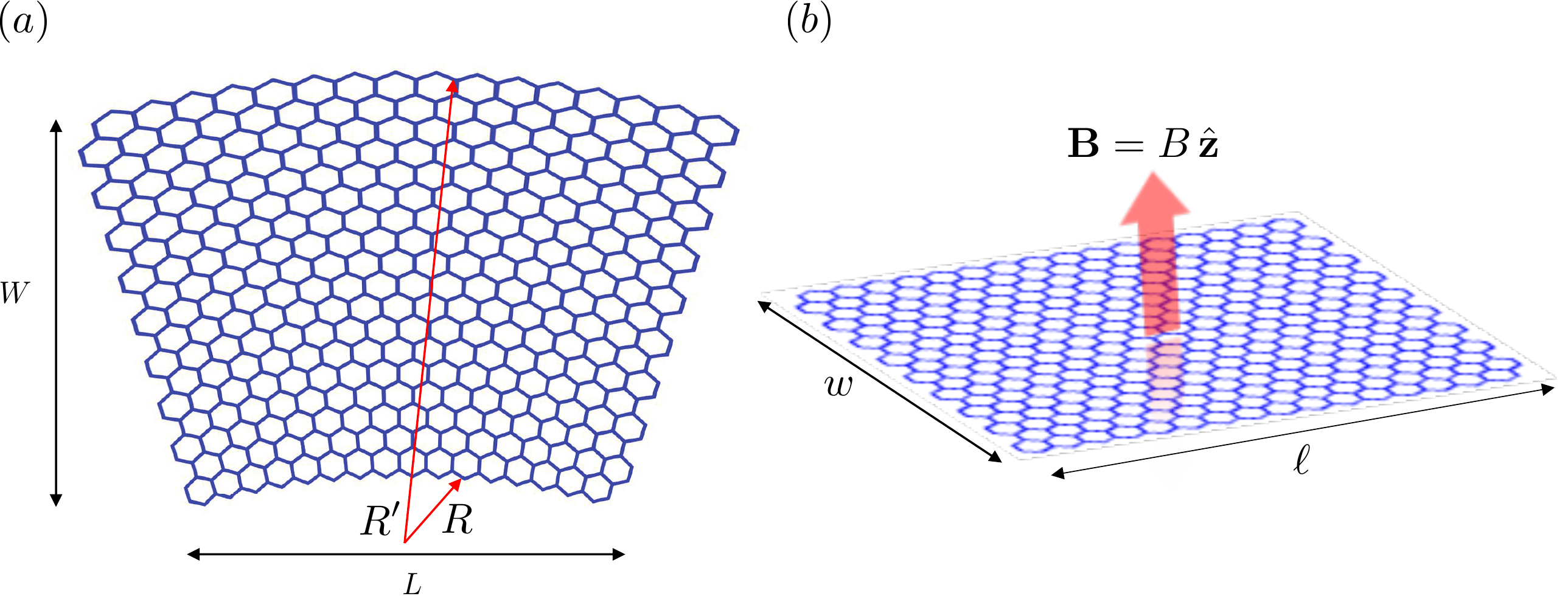}
\caption{(a) Pictorial representation of a rectangular graphene flake deformed into an arc. The radii of the lower and upper edge are, respectively, $R$, and $R'$.  With a graphene flake of width $W=200$ nm, and length $L=192$ nm, and an inner and outer radius of, respectively, $R=5L=960$ nm, and $R'=R+W=1.16$ $\mu$m, the maximum achievable magnitude of the pseudomagnetic field in the central region of the flake is $B=10$ T \cite{ref39}. (b) Flattened equivalent geometry for the bent graphene flake in panel (a). The curvature induced by the bending is replaced with an artificial gauge field, which gives rise to a uniform pseudomagnetic field $\vett{B}=B\,\uvett{z}$ parallel to the $z$-axis. The width $w$ and length $\ell$ of the flattened flake might be extended to infinity, without changing the essential role of the pseudomagnetic field. in the interaction dynamics of the flake with an external electromagnetic pulse.}
\label{figure1}
\end{center}
\end{figure}
where $a\simeq 1.42$ \si{\angstrom} is the carbon-carbon interatomic distance, $\beta=\simeq 2$ is the electron Gr\"uneisen parameter \cite{ref47}, $s$ accounts for the strength of the strain, and $u_{\mu\nu}=\left(\partial_{\mu}u_{\nu}+\partial_{\nu}u_{\mu}\right)$ is the strain tensor, with $u_{\mu}$ being the displacement vector. Notice, that strain also induces a scalar potential $V(x,y)\propto u_{xx}+u_{yy}$, which can be anyway neglected, by choosing an appropriate gauge, where the scalar potential is set to zero. A careful choice of the strain tensor can lead to different bending and deforming geometries, corresponding to different AGFs. Amongst the various choices available \cite{ref46}, we choose the bending profile discussed in Ref. \cite{ref39}, which allows the creation of a uniform pseudomagnetic field in a rectangular graphene flake by introducing only one deformation parameter, i.e., the bending radius $R$. The explicit expression of the displacement vector that implements this geometry is then given as
\bseq
\begin{align}
u_x(x,y) &= \frac{xy}{R},\\
u_{y}(x,y)&=-\frac{x^2}{2R},
\end{align}
\eseq
which corresponds to an AGF $\vett{A}^{(s)}=\pm s\beta y/aR\uvett{x}\equiv-By\uvett{x}$, and, therefore, to a pseudomagnetic field $\vett{B}=B\uvett{z}$. A schematic representation of the bent graphene flake, together with a set of experimentally realisable parameters, is given in Fig. \ref{figure1}. Notice, that for the choice of parameters as in Fig. \ref{figure1}, a uniform magnetic field of magnitude $B=10$ T can be generated within the flake. In general, however, since the magnitude of the pseudomagnetic field is inversely proportional to the bending radius of the graphene flake, i.e., $B=s\beta/R$, a smaller bending radius, i.e., a bigger bending angle, will result in a higher pseudomagnetic field. It is also worth noticing, that introducing a strain also has two other effects, namely it shifts the position of the Dirac points in $k$ space by a quantity proportional to $\vett{A}^{(s)}$ \cite{ref47}, and it also renders Fermi velocity anisotropic \cite{ref48}, according to the relation $\vett{v}_F=v_{F0}\left(\vett{I}-\beta\bar{\vett{u}}+\bar{\vett{u}}\right)$, where $\bar{\vett{u}}$ is the strain tensor. However, for the purpose of this work, we assume that, at the leading order in the bending radius $R$, the Fermi velocity is not affected by the deformation and it remains approximately constant. 

The dynamics of electrons in graphene in the presence of the AGF $\vett{A}^{(s)}$ can be studied by replacing the kinetic momentum $\vett{p}$ in the low-energy graphene Hamiltonian \cite{ref47} $\mathcal{H}=v_F\bm{\sigma}\cdot\vett{p}$ (with $\bm{\sigma}=\sigma_x\uvett{x}+\sigma_y\uvett{y}$, and $\sigma_{\mu}$ are Pauli matrices), with the canonical momentum $\bm{\Pi}=\vett{p}+e\vett{A}^{(s)}$ deriving from minimal coupling of the electron field with the AGF \cite{ref2}, which results in the following Dirac equation for electrons in the vicinity of the Dirac point
\beq\label{dirac0}
i\hbar\partial_t\phi=-i\hbar v_F\sigma^j\left(\partial_{j}+\frac{ie}{\hbar}A_{j}^{(s)}\right)\phi
\eeq
where $j=\{x,y\}$. The above equation can be solved analytically, as it represents the well-known problem of a relativistic electron in a magnetic field, whose solution can be cast in terms of the Landau eigenfunctions
\beq\label{landau1}
\phi_n^{\pm}(y,t,k)=\mathcal{N}\,e^{i\left(kx-\frac{\mathcal{E}_n}{\hbar}t\right)}\left(\begin{array}{c}
\text{sign}\left(n\right)\phi_{|n|-1}(\xi),\\
\\
\phi_{|n|}(\xi)
\end{array}\right),
\eeq
\begin{figure}[t!]
\begin{center}
\includegraphics[width=0.5\textwidth]{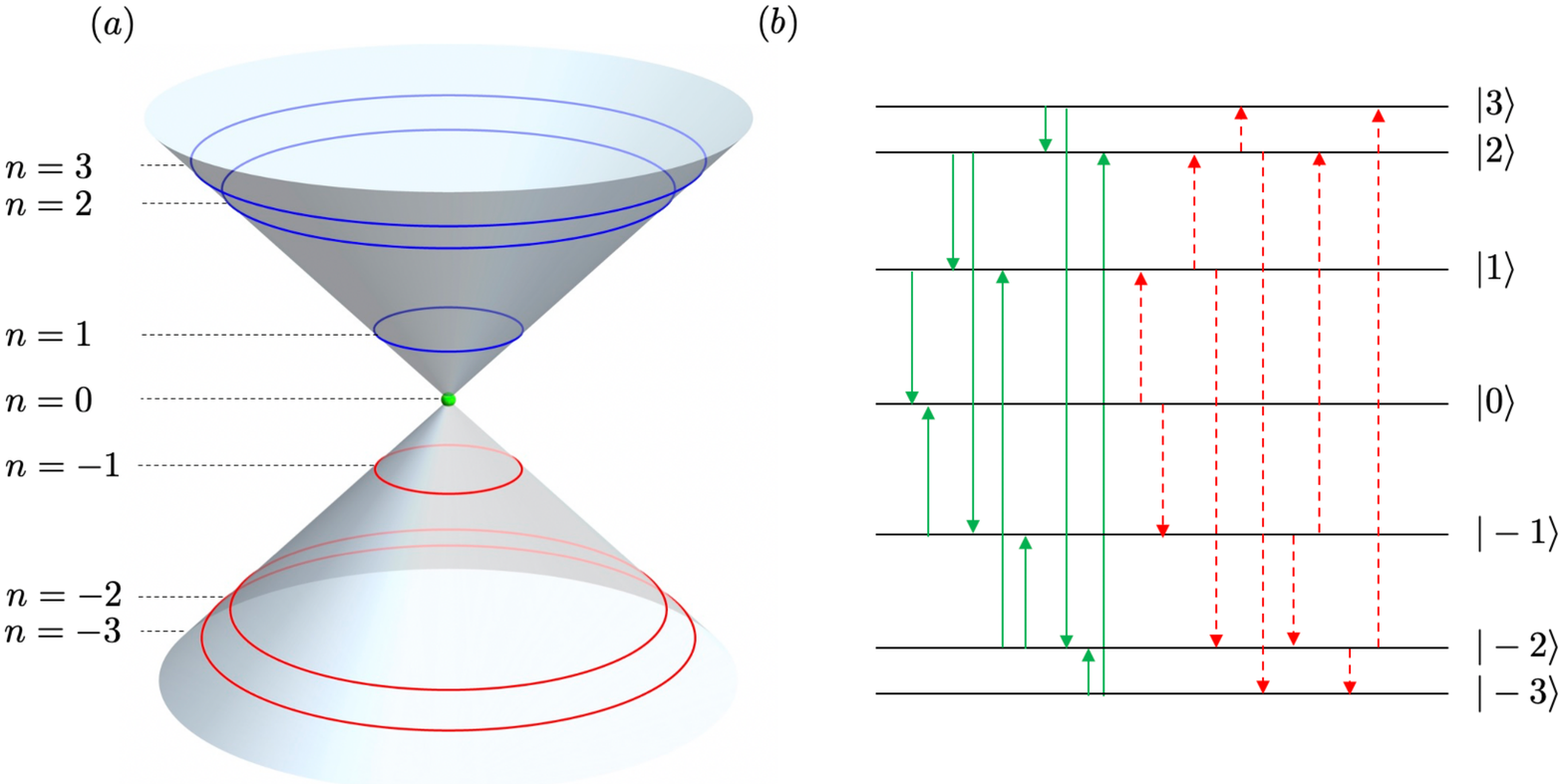}
\caption{(a) Pictorial representation of the band structure of graphene in the vicinity of a Dirac point (solid cones). The presence of a pseudomagnetic field generates Landau levels in both the valence band (red lines) and conduction band (blue lines), as well as at the Dirac point itself (green point). The energies of the Landau levels created in this manner are the same in modulus for the two bands, and differ only in sign, i.e., positive energies are associated with the conduction band, while negative energies with the valence band. (b) Schematic representation of the level structure induced by the pseudomagnetic field, and the correspondent selection rules for the first 3 Landau levels in valence and conduction band. The green, solid arrows correspond to the transition allowed for incoming left-handed circular polarisation (LHC) photons, while the red, dashed arrow correspond to transitions allowed for incoming right-handed circular polarisation (RHC) photons.}
\label{figure2}
\end{center}
\end{figure}
where $\mathcal{N}$ is a normalisation constant, which equals one for $n=0$, and $1/\sqrt{2}$ otherwise,  $\phi_n(\xi)$ are harmonic oscillator eigenstates, $\xi=\left(y+L_c^2k\right)/L_c$,  with $L_c=\sqrt{\hbar/eB}$ being the magnetic length (for a pseudomagnetic field of $B=10$ T we get $L_c\simeq 8.11$ nm), and $\mathcal{E}_n=\text{sign}\left(n\right)\hbar\omega_C\sqrt{|n|}$, with $\omega_C=v_F/L_c$ being the cyclotron frequency. The above solution is moreover equipped with the constraint $\phi_{-1}(\xi)=0$ \cite{ref47}. A sketch of the structure of Landau levels in graphene is depicted in Fig. \ref{figure2}. Notice, that contrary to the case of a non-relativistic electron in magnetic field, where the spacing of Landau levels is constant, i.e., Landau levels have the full structure of a harmonic oscillator, in graphene they are not equally spaced, as the energy eigenvalue $\mathcal{E}_n$ scales as $\sqrt{|n|}$. Moreover, that due to the peculiar band structure of graphene in the vicinity of the Dirac point, two set of Landau levels are created by the magnetic field, corresponding to Landau levels in the valence band [associated to negative values of the index $n$, and appearing as red lines in Fig. \ref{figure2} (a) ], and in the conduction band [associated to positive values of the index $n$, and appearing as blue lines in Fig. \ref{figure2} (a)]. The two set of oscillator states are almost disjoint from each other, with the exception of the ground state $n=0$ [green dot in Fig. \ref{figure2} (a)], which sits at exactly the Dirac point and it is common to both sets.

The creation of Landau levels in graphene introduces selection rules for the dipole-allowed transitions, i.e., an incoming photon can only excite the transition $\ket{n_i}\rightarrow\ket{n_f}$ if and only if $|n_f|=|n_i|\pm 1$, where the plus sign holds for right-handed circularly polarised photons, while the minus sign holds for left-handed circularly polarised photons only. These selection rules are a consequence of the natural spin-orbit coupling arising from the interaction of graphene's band structure with the pseudomagnetic field \cite{ref49}. The structure of these selection rules is shown in Fig. \ref{figure2} (b) for the first three Landau levels in both conduction and valence band. 

\section{Interaction with an Ultrashort Electromagnetic Pulse}\label{section3}
To reach our goal, i.e., investigate the nonlinear response of graphene to an external ultrashort pulse, we first need to understand how to describe its interaction with an external, time-dependent field. To this aim, let us assume that a flake of bent graphene is interacting with an external field within the electric dipole interaction approximation. In this case, therefore, the equation of motion for an electron in the vicinity of a Dirac point reads as follows
\beq\label{eq5}
i\hbar\partial_t\psi=\left[-i\hbar v_F\sigma^{j}\left(\partial_{j}+\frac{ie}{\hbar}A^{(s)}_{j}\right)+e\,E_{j}(t)x^{j}\right].
\eeq
Without any loss of generality, we can assume, that the impinging electric field is linearly polarised along the $x$-direction, and that is characterised by a central frequency $\omega_L$. We also assume, that the electric field is normally impinging on the graphene flake, i.e., that the field is propagating along the $z$ direction, as defined in Fig. \ref{figure1}. We then work in the Landau gauge, where the scalar potential is zero. In this case, we can entirely describe the electric field by means of its vector potential $A(t)=-\int\,dt\,E(t)\equiv\mathcal{A}(t)\exp{[-i\omega_L t]}$, where $\mathcal{A}(t)$ accounts for the temporal shape of the pulse, so that the total vector potential, given by the combination of the true, $A(t)$, and the artificial, $A^{(s)}(y)$ gauge field, can be written as $\vett{A}(y,t)=\left[A(t)-By\right]\uvett{x}$. In this gauge, Dirac equation reduces to the following, manifestly covariant, compact form
\beq\label{dirac1}
i\hbar\gamma^{\mu}\left(\partial_{\mu}+\frac{ie}{\hbar}A_{\mu}\right)\psi=0,
\eeq
where $\mu=\{0,1,2\}\rightarrow\{v_F t,x,y\}$, $\gamma^0=\sigma^z$, $\gamma^1=i\sigma^y$, and $\gamma^2=-i\sigma^x$. In general, this equation does not admit an explicit analytical solution with an arbitrary time-dependent vector potential. However, following the procedure described in Ref. \cite{ref35bis}, if we know the form of the instantaneous eigenstates for the above equation, we can then make an educated guess at the true form of the solution for any time $t$. To this aim, we first take the Fourier transform with respect to the $x$-variable and operate the following phase transformation, i.e.,
\begin{figure*}[!t]
\begin{center}
\includegraphics[width=\textwidth]{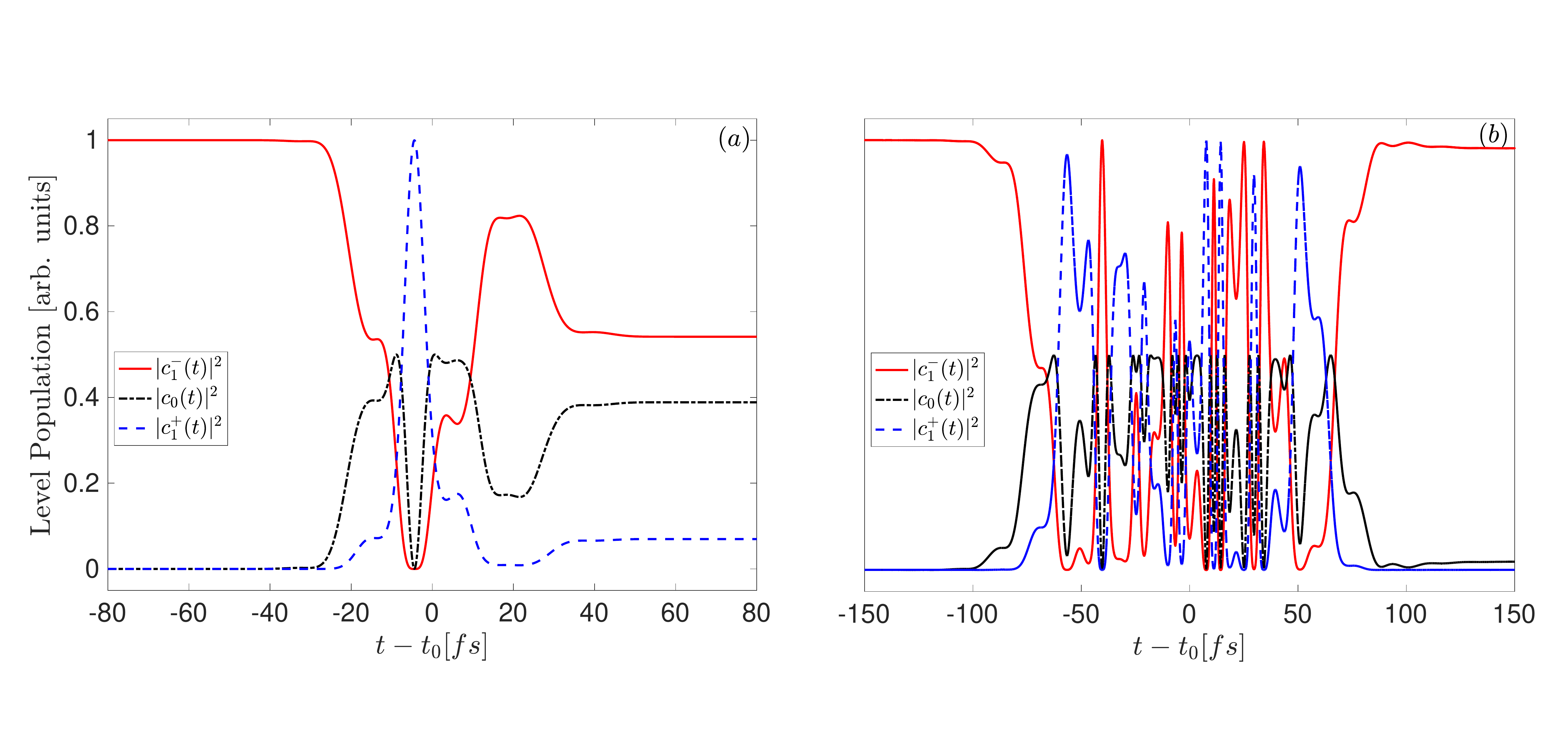}
\caption{Temporal evolution of the population of the Landau levels $\ket{0}$ (black, dot-dahsed line), $\ket{1}$ (blue, dashed line), and $\ket{-1}$ (red, solid line), for different values of the time duration $\tau$ of the impinging ultrashort pulse, i.e., (a) $\tau=20$ fs, and (b) $\tau=50$ fs. The central frequency of the impinging electric field is $\omega_L=\omega_1=174$ THz, and the pulse amplitude is $E_0=10^7$ $V/m$. As it can be seen, as $\tau$ increases, the rate of population transfer also changes accordingly (longer pulses correspond to a higher Rabi frequency \cite{ref51}). However, for the selected configuration, the interaction between levels $\ket{1}$ and $\ket{-1}$ is always mediated by the ground state $\ket{0}$ at the Dirac point. To obtain these plots, the value $v_F=c/300$ m/s for the Fermi velocity has been assumed.}
\label{figure3}
\end{center}
\end{figure*}
\beq
\psi(x,y,t)=\int\,dk\,e^{ikx}e^{iG(t)\sigma_x}\phi(y,t,k),
\eeq
with 
\beq
G(t)=\frac{ev_F}{\hbar}\int_0^t\,d\tau\,A(\tau), 
\eeq
and notice that Eq. \eqref{dirac1} reduces to
\beq
\left[\frac{1}{v_F}\gamma^0\partial_t+\gamma^1\left(-ik-\frac{ieB}{\hbar}y\right)+\partial_y\right]\phi=0,
\eeq
which is equivalent to Eq. \eqref{dirac0}. This indicates, that the Landau eigenstates in Eq. \eqref{landau1} can be taken as the instantaneous eigenstates for the problem at-hand, and that the general solution of Eq. \eqref{dirac1} can be written as
\beq\label{ansatz}
\psi(y,t,k)=\sum_ne^{iG(t)\sigma_x}\left[c_n^+(t)\phi_n^+(y,k)+c_n^-(t)\phi_n^-(y,k)\right].
\eeq
Substituting the above Ansatz into Eq. \eqref{dirac1} leads to a set of coupled mode equations for the expansion coefficients $c_n^{\pm}(t)$, which, for a linearly polarised impinging electromagnetic field read
\bseq\label{coupled1}
\begin{align}
\dot{c}_m^+ &= i\left|\mathcal{N}\right|^2\Omega (t) e^{-i\omega_Lt}\Big[f_{m,m-1}^-(t)c_{m-1}^++f_{m,m+1}^-(t)c_{m+1}^+\nonumber\\
&+\sum_{n<0}f^+_{m,n}(t)\expectation{\phi_m^-}{\sigma^x}{\phi_n^+}c_n^-\Big],\\
\dot{c}_m^- &= - i\left|\mathcal{N}\right|^2\Omega (t) e^{-i\omega_Lt}\Big[f_{m,m-1}^{-,*}(t)c_{m-1}^++f_{m,m+1}^{-,*}(t)c_{m+1}^+\nonumber\\
&+\sum_{n<0}f^{+,*}_{m,n}(t)\expectation{\phi_m^+}{\sigma^x}{\phi_n^-}c_n^-\Big],
\end{align}
\eseq
where $\Omega(t)=(ev_F/\hbar)\mathcal{A}(t)$ is the Rabi frequency, $f^{\pm}_{m,n}(t)=e^{i(\omega_m\pm\omega_n)t}$ accounts for the eigenvalue mismatch between the states participating in the temporal evolution of $c_m^{\pm}(t)$, $\omega_m=\omega_C\sqrt{|m|}$ is the eigenfrequency associated to the state $\ket{\phi_{m}^{\pm}}$ (notice, that the sign of $\omega_m$ has been explicitly taken care of during the calculations that lead to the above equations already), and $\expectation{\phi_m^-}{\sigma^x}{\phi_n^+}$ ($\expectation{\phi_m^+}{\sigma^x}{\phi_n^-}$) is the dipole matrix element of graphene \cite{ref49}, whose explicit expression in this case is given by
\beq
\expectation{\phi_m^-}{\sigma^x}{\phi_n^+}=\delta_{|n|,|m|-1}+\delta_{|n|,|m|+1}.
\eeq

 These equations are the first result of our work. They, in fact, describe the interaction of an arbitrarily shaped linearly polarised impinging field. Notice, how the polarisation of the field enters only in the dipole matrix element, and therefore the above equation can be easily generalised for an arbitrary polarisation, by replacing $\sigma^x$ in the expression of the dipole matrix element, with the Pauli matrix (or combination thereof) correspondent to the impinging polarisation.
 
We can simplify the coupled mode equations above by assuming that the impinging electromagnetic field is nearly resonant with one specific transition. Since the polarisation of the impinging field is linear, rather than circular, both green and red transitions in Fig. \ref{figure2}(b) will be allowed, once the frequency of the incoming field has been chosen to match one of the transitions between the Landau levels. A closer inspection to the structure of the selection rules depicted in Fig. \ref{figure2}(b), moreover, reveals that the only nontrivial dynamics that can be induced in bent graphene corresponds to an impinging field resonant with the transition $\ket{0}\leftrightarrow\ket{\pm 1}$, and a consequent three-level intra-band dynamics, rather than a simple two-level one, either inter-band or intra-band.

If we then now assume, that the impinging field is resonant with the $\ket{0}\leftrightarrow\ket{\pm 1}$ transitions, i.e., $\omega_L=\omega_1=\omega_C$, Eqs. \eqref{coupled1} become
\beq
i\frac{d}{dt}\left(\begin{array}{c}
c_1^-\\
c_0\\
c_1^+
\end{array}\right)=\left(\begin{array}{ccc}
0 & \Gamma(t) & 0\\
\Gamma^*(t) & 0 & -\Gamma(t)\\
0 & -\Gamma^*(t) &0
\end{array}\right)\left(\begin{array}{c}
c_1^-\\
c_0\\
c_1^+
\end{array}\right).
\eeq
where $\Gamma(t)=\Omega(t)\exp{\left(-i\,\omega_1\,t\right)}$. Notice, that the structure of the coefficient matrix in the equations above admits the existence of a dark state, corresponding in this case to no population being at the Dirac point at any given time, i.e., $c_0(t)=0$. The consequences of this, and its benefits for coherent population control of graphene in the presence of an external magnetic field have been recently investigated in Ref. \cite{ref44}.
\begin{figure*}[!t]
\begin{center}
\includegraphics[width=\textwidth]{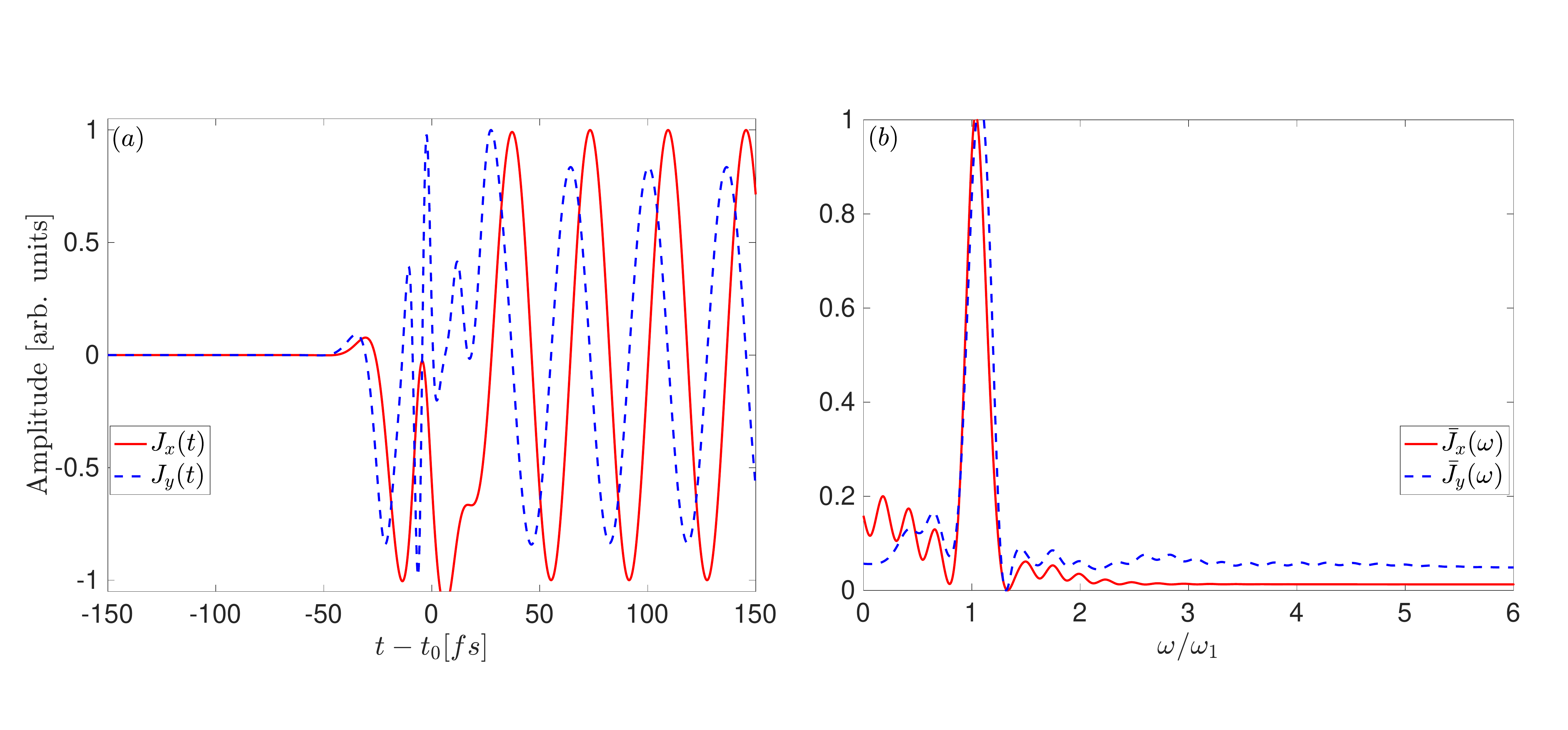}
\caption{Temporal evolution (a) and Fourier transform (b) of the components of the Dirac current $J_x(t)$ (red, solid line), and $J_y(t)$ (blue, dashed line), for an impinging pulse of duration $\tau=20$ fs. As it can be seen from panel (a), despite the fact the impinging electric field is polarised along the $x$-direction, a nonzero current is also generated in the $y$-direction. In particular, panel (b) reveals how while $\tilde{J}_y(\omega)$ contains essentially only a peak at $\omega=\omega_1$, $\tilde{J}_x(\omega)$ has a richer structure. The extra peaks in $\tilde{J}_y(\omega)$ are localised around the fundamental frequency $\omega_1$, and are the source of half-integer harmonics seen in Fig. \ref{figure5} and \ref{figure6}. For these plots, $\omega_L=\omega_1=174$ THz, $E_0=10^7$ V/m, and $v_F=c/300$ m/s have been used, to reproduce results compatible with Fig. \ref{figure3}.}
\label{figure4}
\end{center}
\end{figure*}
We solve the above equations with the initial condition $c_1^-(0)=1$, i.e., with the electron initially in the valence band, and for a vector potential described by a Gaussian pulse of duration $\tau$, and central frequency $\omega_L=\omega_1$, so that $\Omega(t)$ becomes
\beq
\Omega(t)=\frac{ev_FE_0\tau}{\hbar} e^{-\frac{(t-t_0)^2}{\tau^2}}\cos\left(\omega_L t\right),
\eeq
where $E_0$ is measured in $V/m$ and $t_0$ is an arbitrary temporal delay. The evolution of the expansion coefficients $c_n^{\pm}(t)$ for an impinging pulse of amplitude $E_0=10^7$ $V/m$, and time duration $\tau=10$ fs, corresponding to a nearly single-cycle pulse (i.e., $\omega_L\tau=1.74$), and $\tau=50$ fs, corresponding to a pulse with several optical cycles, i.e., $\omega_L\tau=8.7$, is depicted in Fig. \ref{figure3} (a) and (b), respectively. Notice, how a very short pulse ($
\tau=10$ fs induces an almost instantaneous change in the population of the three interested levels, accompained by very faint Bloch-Siegert oscillations \cite{extra15}, as the interaction is too fast, and the system does not have the time to adapt to it. For longer pulses, as in the case of Fig. \ref{figure3} (b), instead, the population dynamics appears more complicated, but leads, at equilibrium, to a situation in which all population returns back to the initial state $\ket{-1}$. In both cases, however, the dynamics always involves the ground state $\ket{0}$ at the Dirac point.

Our simulations, moreover, have been conducted at $T=0$ K, i.e., the distribution of carriers in the valence and conduction band has not been explicitly taken into account. However, as it has been discussed by one of the authors in previous publications \cite{extra13,extra14}, the impact of temperature on the nonlinear response of graphene is negligible.

\section{Nonlinear Signal and Dirac Current}\label{section4}

The nonlinear response of graphene can be estimated by evaluating, as a function of frequency, the intensity of the nonlinear radiation emitted by it, as a consequence of the interaction with an impinging, time-dependent electric field, i.e. \cite{ref35bis}
\beq\label{nonlinearS}
I(\omega)\propto\left|\omega\,\tilde{\vett{J}}(\omega)\right|^2,
\eeq
where $I(\omega)$ is the spectrum of the emitted radiation, and $\tilde{\vett{J}}(\omega)$ is the Fourier transform of the Dirac current $j^{\mu}(x,y,t)=\bar{\psi}(x,y,t)\gamma^{\mu}\psi(x,y,t)=\psi^{\dagger}(x,y,t)\sigma^{\mu}\psi(x,y,t)$ (where the last equality holds because of our definition of gamma matrices given in the previous section).

Since we are only interested in the temporal features of the current, we can integrate it with respect to the transverse space, to obtain
\beq\label{eq17}
\vett{J}(t)=\int\,d^2R\,\vett{j}(x,y,t)=\int\,d^2R\,\left(\bm{\sigma}\cdot\vett{R}\right)\left|\psi(x,y,t)\right|^2,
\eeq
\begin{figure*}[!t]
\begin{center}
\includegraphics[width=\textwidth]{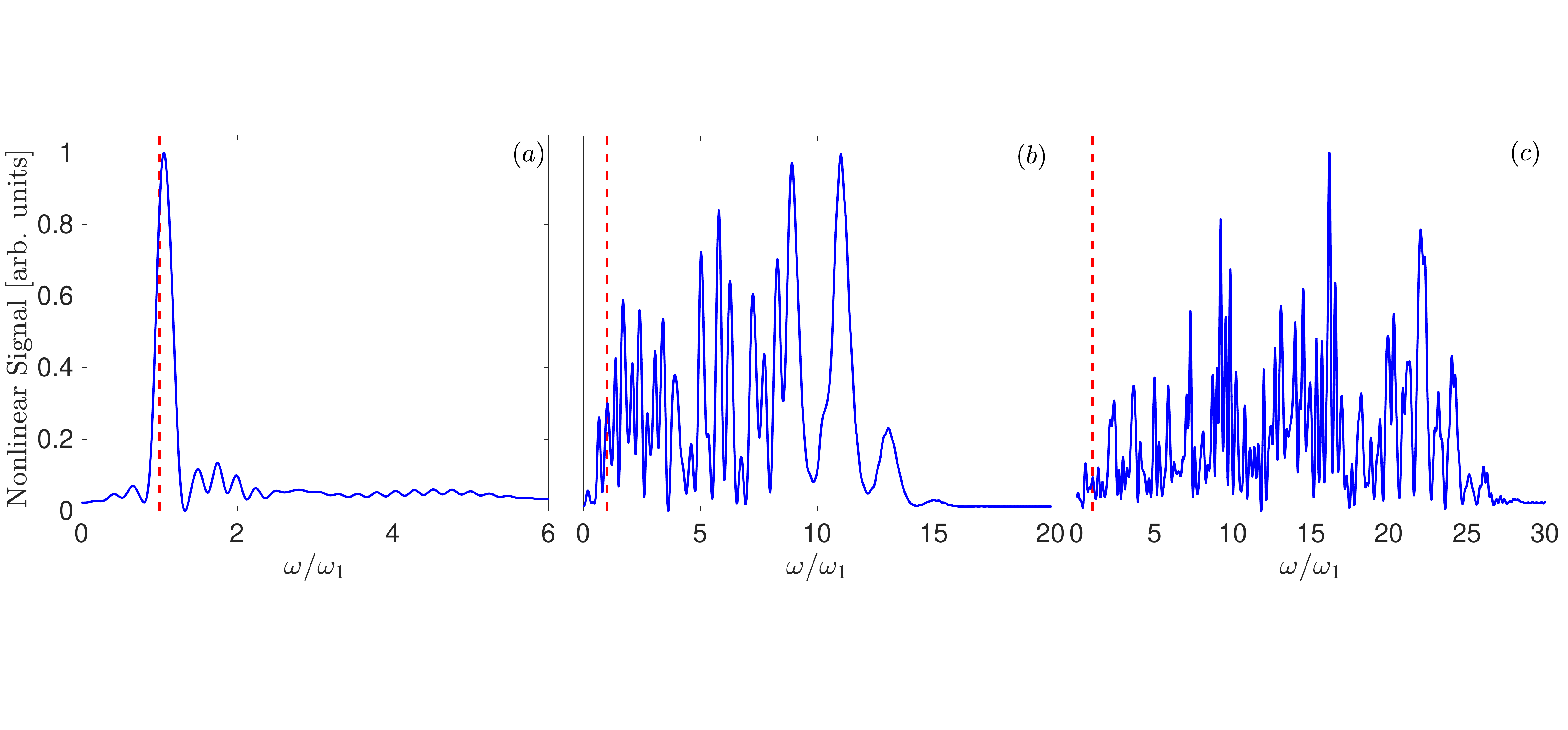}
\caption{Nonlinear signal, as defined in Eq. \eqref{nonlinearS}, as a function of the normalised frequency $\omega/\omega_1$, for different values of the pulse length, i.e., (a) $\tau=20$ fs (correspondent to a fluence of $F=2.65$ $m\,J/m^2$), (b) $\tau=50$ fs (correspondent to a fluence of $F=6.64$ $m\,J/m^2$), and (c) $\tau=100$ fs (correspondent to a fluence of $F=13.27$ $m\,J/m^2$). The red, dashed line in each panel represents the position of the fundamental frequency $\omega_L=\omega_1$. Notice, that the peak that should correspond to $\omega=\omega_1$ is indeed slightly blue-shifted, due to the different weight of the $x$- and $y$-components of the Dirac current, as it can be seen in Fig. \ref{figure4}(b). This effect only manifests significantly for short pulses, and tends to disappear for longer ones, as it can be seen by comparing panel (a) and (c), above. For these plots, $\omega_L=\omega_1=174$ THz, $E_0=10^7$ V/m, and $v_F=c/300$ m/s have been used, to reproduce results compatible with Fig. \ref{figure3}. }
\label{figure5}
\end{center}
\end{figure*}
Notice, that since the expansion coefficients $c_n^{\pm}(t)$ appearing in the definition of $\psi(x,y,t)$ given by Eq. \eqref{ansatz} only depend on time, and that the solution along the $x$-direction can be expressed in terms of plane waves, we can safely perform the integration only over the Landau eigenstates, and exploit their orthogonality relation to compute the spatial integral in Eq. \eqref{eq17}. In doing this, moreover, the resulting time-dependent current $\vett{J}(t)$ will have the same functional form for both the case of a finite graphene flake undergoing bending, and an infinite sheet of graphene subjected to a uniform magnetic field orthogonal to the graphene plane. The only difference between the two cases will then be an overall multiplicative constant, which will account for the actual arrangement of the system. Since this constant does not change the overall functional for of $\vett{J}(t)$, and neither impacts the overall form of its Fourier transform, we then treat it as a simple normalisation factor, and scale everything to it.

If we then substitute Eq. \eqref{ansatz} into the above expression, and limit ourselves to the case of the impinging pulse being resonant with the $\ket{0}\leftrightarrow\ket{\pm 1}$, the $x$- and $y$-components of the integrated current have the following explicit form
\bseq
\begin{align}
J_x(t) &=c_0^*(t)\left[c_1(t)^+e^{-i\omega_1t}-c_1(t)^-e^{i\omega_1t}\right]+\text{c.c.},\\
J_y(t) &=i\,c_0^*(t)\left[c_1(t)^+e^{-i\omega_1t}-c_1(t)^-e^{i\omega_1t}\right]+\text{c.c.}.
\end{align}
\eseq

The temporal evolution of the Dirac current, as well as its Fourier transform,  is shown in Fig. \ref{figure4} for the case of an impinging pulse with $\tau=10$ fs. Notice how, despite the fact the impinging electric field is polarised along the $x$-direction, a nonzero component of the current along the $y$-direction arises. This is a consequence of the broken centrosymmetry, induced by the artificial gauge field.  Since the current enters with its whole vectorial character in the definition of the nonlinear signal as given by Eq. \eqref{nonlinearS}, this has significant consequences on its spectrum. In fact, for the case of a short pulse as the one used in Fig. \ref{figure4}, the shape of the nonlinear signal is almost entirely determined by $\tilde{J}_x(
\omega)$, while for longer pulse widths, the interplay between the current components $J_{\mu}(t)$ becomes more prominent, giving rise to a richer structure of $I(\omega)$. 

To prove this, in Fig. \ref{figure5} we plot the nonlinear signal defined in Eq. \eqref{nonlinearS} for different values of the impinging pulse width, i.e., $\tau=10$ fs [panel (a)], $\tau=50$ fs [panel (b)], and $\tau=100$ fs [panel (c)]. As it can be seen, for very short pulses [panel (a)], a set of equally spaced harmonics appear, with spacing $n\omega_1/2$.This can be explained by noticing, that the impinging pulse effectively sees a 3-level system with equally spaced levels, as $\ket{\pm 1}$ have the same distance in energy from $\ket{0}$. This allows us to approximate the square-root behaviour of the Landau ladder to the more traditional harmonic oscillator behaviour which, in turn, gives the equally spaced peaks in Fig. \ref{figure5}(a), with spacing $\omega_1/2$. 

This is the second result of our work. For short enough pulses, the local structure of Landau levels around the Dirac point can be approximated with that of a traditional harmonic oscillator, and thus the corresponding nonlinear signal contains all the integer and half-integer harmonics of the pulse carrier frequency $\omega_L$.

For longer pulse durations, on the other hand, the nonlinear signal shows a more rich and complicated spectrum. From Fig. \ref{figure5}(b) and (c), in fact, it is possible to see how the spectrum broadens, with respect to the situation depicted in Fig. \ref{figure5}(a), and higher harmonics appear, containing both even and odd contributions. Notice, moreover, how in both cases of Fig. \ref{figure5}(b) and (c), although the highest intensity is reached for the 7th  [panel (b)] and the 11th [panel (c)] harmonic, higher ones still have considerable intensity. For example, for long pulses [panel (c)], the 20th can be generated with significant intensity.

If we recall that $\omega_1=174$ THz, $\omega=20\,\omega_1$ would correspond to radiation in the visible region, with a wavelength of approximately $\lambda_{20th}=540$ nm. Moreover, harmonics as high as the 25th-26th can be also generated (although with small intensity). This corresponds to the blue side of the visible spectrum, as $\lambda_{25th}= 433$. nm. The nonlinear signal in Fig. \ref{figure5}(c), therefore, spans the whole spectrum between THz and the visible region, and could also be pushed into the near UV. This suggests that artificial gauge fields in graphene could be used as a mean to efficiently convert signals within these spectral region.
\section{Effect of the Pseudomagnetic Field on the Nonlinear Signal}
We now briefly discuss what is the effect of the pseudomagnetic field on the nonlinear signal. To do that, we focus our attention on the case of a $\tau=50$ fs pulse, and repeat our simulations using different values of the pseudomagnetic field. This, in practice, would correspond to a larger or smaller bending radius (and, consequently, bending angle) for smaller or greater values of the pseudomagnetic field, respectively (See Fig. \ref{figure1}).

\begin{figure*}[!t]
\begin{center}
\includegraphics[width=\textwidth]{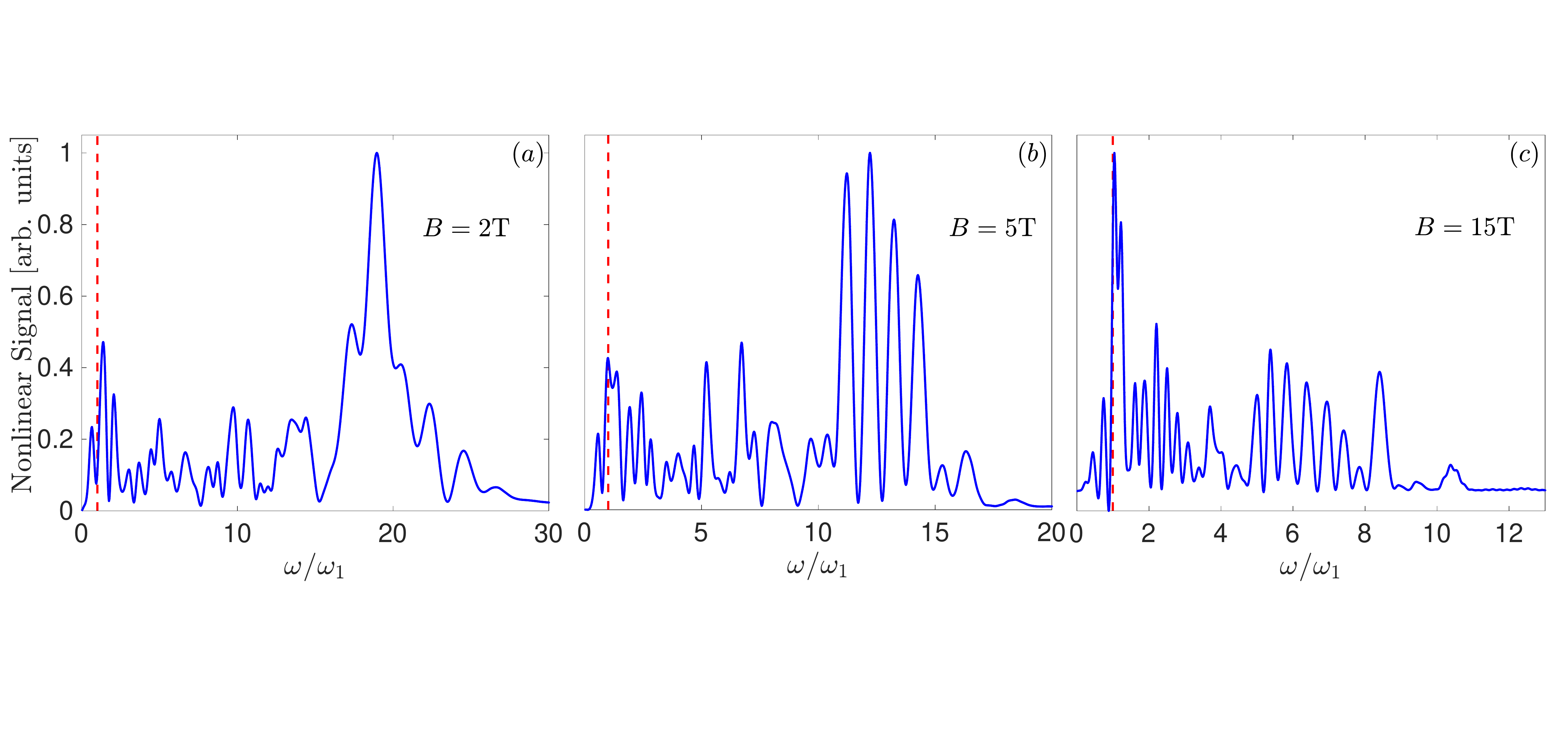}
\caption{Nonlinear signal corresponding to an impinging electromagnetic pulse of duration $\tau=50$ fs, for different values of the pseudomagnetic field, namely (a) $B=2$ T, (b) $B=5$ T, and (c) $B=15$ T. For low pseudomagnetic fields, we observe an almost resonant transfer of energy between the impinging field as its 20th harmonics [panel (a)], or an almost equal distribution of energy of the impinging pulse across its 10th-15th harmonics [panel(b)]. For high values of the pseudomagnetic field [panel (c)], instead, we observe the appearance of half-integer harmonics centreda round $\omega/\omega_1\simeq 2-3$ and $\omega/\omega_1\simeq 6-8$. For these plots, $\omega_L=\omega_1=174$ THz, $E_0=10^7$ V/M (correspondent to a fluence $F=6.64$ $m\,J/m^2$), and $v_F=c/300$ m/s have been used, to reproduce results compatible with Fig. \ref{figure5}.}
\label{figure6}
\end{center}
\end{figure*}

We show the results of these simulations in Fig. \ref{figure6}, where the nonlinear signal il plotted against three different values of the pseudomagnetic field, namely $B=2$ T [Fig. \ref{figure6} (a)], $B=5$ T [Fig. \ref{figure6} (b)], and $B=15$ T [Fig. \ref{figure6} (c)].

As it can be seen, for large values of $B$ [Fig. \ref{figure6} (c)] with respect to those employed in Fig. \ref{figure5}, half-integer harmonics centred around $\omega/\omega_1\simeq 2-4$ and $\omega/\omega_1\simeq 6-8$ distinctively appear. For $B=2$ T, on the other hand, a very efficient transfer of energy between the impinging field at $\omega_L=\omega_1$ and its 20th harmonics takes place, which means that in the presence of small pseudomagnetic fields (i.e., large bending angles), graphene behaves as a very efficient frequency converter between the impinging field oscillating at $\omega_L=174$ THz and its 20th harmonics, that sits well-within the visible region, at about $\lambda=540$ nm.
\section{Conclusion and Outlook}\label{section5}
In conclusion, we have investigated the nonlinear response of a bent graphene flake to an impinging, linearly polarised, time-dependent electromagnetic field. We have shown, that the AGF induced by bending, and the consequent appearance of Landau levels significantly modify the harmonic signal generated by graphene. in particular, we have shown, that for the case of an impinging field resonant with the transition $\ket{0}\leftrightarrow\ket{\pm 1}$, i.e., with the Landau ladder in the vicinity of the Dirac point, the nonlinear signal shows a spectrum containing integer, as well as non-integer harmonics of the pulse carrier frequency. In particular, for ultrashort pulses, the nonlinear signal contains integer harmonics $\omega=n\omega_L$ as well as half-integer harmonics $\omega=(n+1/2)\omega_L$, while as the pulse gets longer, a more complicated scenario arises, with a nonlinear signal containing up to the 26th harmonics in its spectrum. Lastly, we have shown that the pseudomagnetic field induced by the AGF breaks the central symmetry typical of graphene, thus allowing the emission of even harmonics of the impinging pulse carrier frequency. The magnitude of the pseudomagnetic field can be controlled by changing the radius of curvature (or, equivalently, the bending angle) of the graphene flake. By doing so, we observe, for ``small" magnetic fields, the almost resonant transfer of energy between the impinging electromagnetic pulse at $\omega=\omega_L$ and its 20th harmonics. For an impinging pulse centred around $174$ THz, as in our case, the 20th harmonics lies well-within the visible spectrum, around $\lambda=540$ nm. This effect, controllable by controlling the magnitude of the pseudomagnetic field, could potentially pave the way for novel graphene-based photonic devices such as THz-to-visible frequency converters, and frequency generators.

In this manuscript, we have focused our attention on the role AGFs have on the nonlinear optical response of graphene to an ultrashort pulse impinging upon it at zero temperature. However, our model can be readily extended to account for a finite temperature, using the more standard Dirac-Bloch model, which will give us the possibility to include other competing effects, such as dephasing, relaxation dynamics, and Coulomb interactions. 

In this work, we have not explored the role of the valley degree of freedom on the nonlinear response of strained graphene. It is known, however, that introducing such a strain can break the valley symmetry \cite{extra1, extra2}. This could then possibly lead to a valley-contrasting nonlinear optical response, which could have significant impact on valleytronic applications. To account for this, one would need to explicitly take spin-orbit coupling into account, such that the combination of AGF and spin-orbit coupling will lift the pseudospin degeneracy in both bulk and edge states, leading to time reversal symmetry breaking \cite{extra3} and, ultimately, to a valley-dependent nonlinear response.

The interplay between AGFs and spin-orbit coupling and their effect on the nonlinear response of strained graphene, together with the extension of our analysis to other 2D materials, such as transition metal dichalcogenides (where we can directly probe the effect of Coulomb interactions on their nonlinear response in the presence of AGFs), and the extension of our formalism to 3D Dirac materials, such as topological semimetals \cite{extra4, extra5, extra6}, will be the subject of our next research.
\begin{acknowledgements}
M.O. acknowledges the support of the Academy of Finland Flagship Programme, Photonics research and Innovation (PREIN), decision 320165.
\end{acknowledgements}
\section*{Data Availability}
The data that support the findings of this study are available from the corresponding author upon reasonable request.
\nocite{*}
\section*{References}
%\bibliography{aipsamp}% Produces the bibliography via BibTeX.

\begin{thebibliography}{99}

%
\bibitem{ref1} J.D. Jackson, ``Classical Electrodynamics", 3rd Edition (Wiley, New York, 1998).
%
\bibitem{ref2} M. D. Schwartz, ``Quantum Field Theory and the Standard Model" (Cambridge university Press, Cambridge, 2013).
%
\bibitem{ref3}Y. Aharonov, and D. Bohm, ``Significance of Electromagnetic Potentials in Quantum Theory", Phys. Rev. \textbf{115}, 485 (1959).
%
\bibitem{ref4}T. T. Wu, and C. N. Yang, ``Concept of nonintegrable phase factors and global formulation of gauge fields", Phys. Rev. D \textbf{12}  ,3845 (1975).
%
\bibitem{ref5}B. Simon, ``Holonomy, the quantum adiabatic theorem, and berry’s phase", Phys. Rev. Lett. \textbf{51}, 2167 (1983)
%
\bibitem{ref6}M. Nakahara, ``Geometry, topology and physics", (CRC Press, Boca Raton, 2003).
%
\bibitem{ref7}M. V. Berry, "Quantal Phase Factors Accompanying Adiabatic Changes". Proc. R. Soc. Lond. A. \textbf{392}, 45 (1984).
%
\bibitem{ref8}S. Pancharatnam, ``Generalized theory of interference and its applications", in ``Proceedings of the Indian Academy of Sciences-Section A", Vol. 44, pp.398-417  (Springer, Berlin, 1956).
%
\bibitem{ref9}A. A. Abrikosov, ``The magnetic properties of superconducting alloys", J. Phys. Chem. Solids. \textbf{2}, 199 (1957).
%
\bibitem{ref10}D. J. Thouless, M. Kohmoto, M. P. Nightingale, and M. den Nijs, ``Quantized hall conductance in a two-dimensional periodic potential", Phys. Rev. Lett. \textbf{49}, 405 (1982).
%
\bibitem{ref11}M. Kohmoto, ``Zero modes and the quantized hall conductance of the two-dimensional lattice in a magnetic field",Phys. Rev. B \textbf{39}, 11943 (1989).
%
\bibitem{ref13}E. J. Yarmchuk, M. J. V. Gordon, and R. E. Packard, ``Observation of Stationary Vortex Arrays in Rotating Superfluid Helium", Phys. Rev. Lett. \textbf{43}, 214 (1979).
%
\bibitem{ref14}M. Ornigotti, G. Della Valle, D. Gatti, and S. Longhi, ``Topological suppression of optical tunneling in a twisted annular fiber", Phys. Rev. A \textbf{76}, 023833 (2007).
%
\bibitem{ref15}S. D. Huber, ``Topological Mechanics", Nat. Phys. \textbf{12}, 621 (2016).
%
\bibitem{ref16}B. A. Bernevig, and T. L. Hughes, ``Topological insulators and topological superconductors", (Princeton University Press,Princeton,, 2013).
%
\bibitem{ref17}N. Goldman, J. C. Budich, and P. Zoller, ``Topological quantum matter with ultracold gases in optical lattices", Nat. Phys. \textbf{12}, 639 (2016).
%
\bibitem{ref18}J. Dalibard, F. Gerbier, G. Juzeliu\={n}as, and P. \"{O}hberg, ``Colloquium: Artificial gauge potentials for neutral atoms", Rev. Mod. Phys. \textbf{83}, 1523 (2011).
%
\bibitem{ref19}N. Goldman, G. Juzeliu\={n}as, P.\"{O}hberg, and I. B. Spielman, ``Light-induced gauge fields for ultracold atoms", Rep. Prog.  Phys. \textbf{77}, 126401 (2014).
%
\bibitem{ref20}J. Dalibard, ``Introduction to the physics of artificial gauge fields", arXiv:1504.05520 (2015).
%
\bibitem{ref21}L. Lu, J. D. Joannopoulos, and M. Solja\v{c}i\'{c},``Topological Photonics", Nat. Photon. \textbf{8}, 821 (2014).
%
\bibitem{ref22}T. Ozawa, H. M. Price, A. Amo, N. Goldman, M. Hafezi, L. Lu, M. C. Rechtsman, D. Schuster, J. Simon, O. Zilberberg, and I. Carusotto, ``Topological Photonics", Rev. Mod. Phys. \textbf{91}, 015006 (2019).	
%
\bibitem{ref23}H. J. Rothe, ``Lattice gauge theories", 4th Edition (World Scientific, Singapore, 2012).
%
\bibitem{ref24}M. C. Rechtsmann, J. Zeuner, Y. Plotnik, Y. Lumer, D. Podolsky, F. Dreisow, S. Nolte, M. Segev, and A. Szameit, ``Photonic Floquet topological insulator", Nature \textbf{496}, 196 (2013).
%
\bibitem{ref24bis}L. J. Maczewsky, M. Heinrich, M. kremer, S. K. Ivanov, M. Ehrhardt, F. Martinez, Y. V. Kartashov, V. V. Konotp, L. Torner, D. bauer, and A. Szmaiet, ``Nonlinearity-induced photonic topological insulator", Science \textbf{370}, 701 (2020).
%
\bibitem{ref25}A. B. Khanikaev, S. H. Mousavi, W.-K. Tse, M. Kargarian, A. H. MacDonald, and G. Shvets, ``Photonic topological insulators", Nat. Mater. \textbf{12}, 233 (2013).
%
\bibitem{ref26}M. Kremer, L. Teuber, A. Szameit, and S. Scheel, ``Optimal design strategy for non-Abelian geometric phases using Abelian gauge fields based on quantum metric", Phys. Rev. Research \textbf{1}, 033117 (2019).
%
\bibitem{ref27}C. Jörg, G. Queraltó, M. Kremer, G. Pelegrí, J. Schulz, A. Szameit, G. von Freymann, J. Mompart, and V. Ahufinger, ``Artificial gauge field switching using orbital angular momentum modes in optical waveguides", Light: Science \& Applications \textbf{9}, 150 (2020).
%
\bibitem{ref28} M. C. Rechtsman, J. M. Zeuner, A. Tünnermann, S. Nolte, M. Segev, and  A. Szameit, ``Strain-induced pseudomagnetic field and photonic Landau levels in dielectric structures", Nat. Phot. \textbf{7}, 153 (2013).
%
\bibitem{ref29}M. C. Rechtsman, Y. Lumer, Y. Plotnik, A. Perez-Leija, A. Szameit, and M. Segev, ``Topological protection of photonic path entanglement", Optica \textbf{3}, 925 (2016).
%
\bibitem{ref30}J. Noh, W. A. Benalcazar, S. Huang, M. J. Collins, K. P. Chen, T. L. Hughes, and M. C. Rechtsman, ``Topological protection of photonic mid-gap defect modes", Nat. Phot. \textbf{12}, 408 (2018).
%
\bibitem{ref31}M. Wang, C. Doyle, B. Bell, M. J. Collins, E. Magi, B. J. Eggleton, M. Segev, and A. Blanco-Redondo, ``Topologically protected entangled photonic states", Nanophot. \textbf{8}, 1327 (2019).
%
\bibitem{ref32}K. S. Novoselov, A. K. Geim, S. V. Morozov, D. Jiang, Y. Zhang, S. V. Dubonos, I. V. Grigorieva, and A. A. Firsov, ``Electric Field Effect in Atomically Thin Carbon Films", Science, \textbf{306}, 666 (2004).
%
\bibitem{ref33}K. S. Novoselov, A. K. Geim, S. V. Morozov, D. Jiang, M. I. Katsnelson, I. V. Grigorieva, S. V. Dubonos, and A. A. Firsov, ``Two-dimensional gas of massless Dirac fermions in graphene", Nature \textbf{438}, 197 (2005).
%
\bibitem{ref34}Yuanbo Zhang, Yan-Wen Tan, Horst L. Stormer, and P. Kim, ``Experimental observation of the quantum Hall effect and Berry's phase in graphene", Nature \textbf{438}, 201 (2005).
%
\bibitem{ref35}M. I. Katsnelson, ``Zitterbewegung, chirality, and minimal conductivity in graphene,” Eur. Phys. J. B.\textbf{51}, 157
(2006).
%
\bibitem{ref35a}R.R.Nair, P.Blake, A.N.Grigorenko, K.S.Novoselov, T.J.Booth, T.Stauber, N.M.R.Peres, and A.K.Geim,
“Fine Structure Constant Defines Visual Transparency of Graphene,” Science \textbf{320}, 1308 (2008).
%
%\bibitem{ref35ter}
%
\bibitem{ref36}M.A.H. Vozmediano, M.I. Katsnelson, and F. Guinea, ``Gauge fields in graphene", Phys. Rep. \textbf{496} 109 (2010).
%
\bibitem{ref37}H. Suzuura, and T. Ando, ``Phonons and electron-phonon scattering in carbon nanotubes", Phys. Rev. B \textbf{65}, 235412 (2002).
%
\bibitem{ref38}J. L. Ma$\tilde{\text{n}}$es, ``Symmetry-based approach to electron-phonon interactions in graphene", Phys. rev. B \textbf{76}, 045430 (2007).
%
\bibitem{ref39}F. Guinea, A. K. Geim, M. I. Katsnelson, and K. S. Novoselov, ``Generating quantizing pseudomagnetic fields by bending graphene ribbons", Phys. Rev. B \textbf{81}, 035408 (2010).
%
\bibitem{ref40}F. Guinea, M. I. Katsnelson, and M. A. H. Vozmediano, ``Midgap states and charge inhomogeneities in corrugated graphene", Phys. Rev. B \textbf{77}, 075422 (2008).
%
\bibitem{ref41}A. Cortijo, and M. A. H. Vozmediano, ``Electronic properties of curved graphene sheets", EPL \textbf{77}, 47002 (2007).
%
\bibitem{extra7} A. R. Wright, X. G. Xu, J. C. Cao, and Z. Chang, ``Strong nonlinear response in the terahertz regime", Appl. Phys. Lett. \textbf{95}, 072101 (2009).
%
\bibitem{extra8} E. Hendry, P. J. Hale, J. Moger, and A. K. Savchenko, ``Coherent nonlinear optical response of graphene", Phys. Rev. Lett. \textbf{105}, 097401 (2010).
%
\bibitem{extra9} H. A. Hafez, S. Kovalev, K.-J. Tielrooij, M. Bonn, M. Gensch, and D. Turchinovich, ``The nonlinear optics of graphene: from saturable absorption to high-harmonic generation", Adv. opt. Mater. \textbf{8}, 1900771 (2020).
%
\bibitem{extra10} G. Demetriou, H. T. Bookey, F. Biancalana, E. Abraham, Y. Wang, W. Ji, and A. K. Kor, ``Nonlinear optical properties of multilayer graphene in the infrared", Opt. Express \textbf{24}, 13033 (2016).
%
\bibitem{extra17} J. B. Khurgin, ``Graphene - A rather ordinary nonlinear optical material", Appl. Phys. Lett. \textbf{104}, 161116 (2014).
%
\bibitem{extra11}R. W. Boyd, ``Nonlinear Optics", 3rd Edition (Academic Press, Cambridge, MA, 2008).
%
\bibitem{extra12}  X. Yao, and A. Belyonin, ``Nonlinear optics of graphene in a strong magnetic field", J. Phys.: Condens. Matter \textbf{25}, 054203 (2013).
%
\bibitem{ref42}Vinu Lukose, R. Shankar,  and G. Baskaran, ``Novel Electric Field Effects on Landau Levels in Graphene", Phys. Rev. Lett. \textbf{98}, 116802 (2007).
%
\bibitem{ref43}A. L\`{o}pez, A. Di Teodoro, J. Schliemann, B. Berche, and B. Santos, ``Laser-induced modulation of the Landau level structure in single-layer graphene", Phys. Rev. B \textbf{92}, 235411 (2015).
%
\bibitem{ref44}C. Ding, R. Yu, X. Hao, and D. Zhang, ``Controllable population dynamics in Landau-quantized graphene", Sci. rep. \textbf{8},  1530 (2018).
%
\bibitem{ref46}F. Guinea, M. I. Katsnelson, and A. K. Geim, ``Energy gaps and a zero-field quantum Hall effect in graphene by strain engineering", nat. Phys. \textbf{6}, 30 (2010).
%
\bibitem{ref47}M. I. Katsnelson, ``Graphene: Carbon in Two Dimensions", (Cambridge University Press, Cambridge, 2012).
%
\bibitem{ref48}M. Oliva-Leyva, and G. G. Naumis, ``Anisotropic AC conductivity of strained graphene", J. Phys.: Condes. Matter \textbf{26}, 125302 (2014).
%
\bibitem{ref35bis}K. L. Ishikawa, "Nonlinear optical response of graphene in time domain", Phys. Rev. B \textbf{82}, 201402(R) (2010).
%
\bibitem{extra15} M. Wegener, ``Extreme Nonlinear optics" (Springer, Berlin, 2005).
%
\bibitem{extra13} D. N. Carvalho, A. Marini, and F. Biancalana, ``Dynamical centrosymmetry breaking - A novel mechanism for second-harmonic generation in graphene", Ann. Phys. \textbf{378}, 24 (2017).
%
\bibitem{extra14}D. N. Carvalho, F. Biancalana, and A. Marini, ``Monolayer graphene can emit SHG waves", Opt. Data Process. Storage, \textbf{3}, 47 (2017).
%
\bibitem{ref49}D. S. L. Abergel, and V. I. Fal’ko, ``Optical and magneto-optical far-infrared properties of bilayer graphene", Phys. Rev. B \textbf{75}, 155430 (2007).
%
\bibitem{ref51}R. Loudon, ``The Quantum theory of Light", 3rd Edition (Oxford Science Publications, Oxford, 2000).
%
\bibitem{extra1}Y. S. Ang, S. Sultan, and C. Zhang, ``Nonlinear optical spectrum of bilayer graphene in the terahertz regime", Appl. Phys. Lett. \textbf{97}, 243110 (2010).
%
\bibitem{extra2} Y. Jiang, T. Low, K. Chang, M. I. Katsnelson, and F. Guinea, ``Generation of pure valley current in graphene", Phys. Rev. Lett. \textbf{110}, 046601 (2015).
%
\bibitem{extra3}W.-Y. He, and L. He, ``Coupled spin and pseudomagnetic field in graphene nanoribbons", Phys. Rev. B \textbf{28}, 085411 (2013).
%
\bibitem{extra4} K. J. A. Ooi, Y. S. Ang, Q. Zhan, D. T. H. Han, L. K. Ang, and C. K. Ong, ``Nonlinear plasmonics of three-dimensional Dirac semimetals", APL Photonics \textbf{4}, 034402 (2019).
%
\bibitem{extra5} J. Lim, Y. S. Ang, F. J. Garcia de Abajo, I. Kaminer, L. K. Ang, and L. J. Wong, ``Efficient generation of terahertz harmonics in three-dimensional Dirac semimetals", Phys. Rev. Research \textbf{2}, 043252 (2020).
%
\bibitem{extra6}J. Lim, K. J. A. Ooi, C. Zhang, L. K. Ang, and Y. S. Ang, ``Broadband strong optical dichroism in topological Dirac semimetals with Fermi velocity anisotropy", Chinese Phys. B \textbf{29}, 077802 (2020).


\end{thebibliography}

\end{document}